\documentstyle[preprint,aps,eqsecnum]{revtex}

\newcommand{\dop}{\bar{\delta}^{(0)}_\parallel}
\newcommand{\dov}{\bar{\delta}^{(0)}_\perp}
\newcommand{\dlp}{\bar{\delta}^{(1)}_\parallel}
\newcommand{\dlv}{\bar{\delta}^{(1)}_\perp}
\newcommand{\Dlp}{\bar{\Delta}^{(1)}_\parallel}
\newcommand{\Dlv}{\bar{\Delta}^{(1)}_\perp}

\newcommand{\tr}{{\rm Tr}}
\newcommand{\xis}{\xi(\sigma)}
\newcommand{\xip}{\xi(\Pi)}
\newcommand{\del}{\partial}
\newcommand{\deldel}{\tensor{\del}}
\newcommand{\VEV}{\Sigma^{(0)}}

\newcommand{\xl}{\xi_L}
\newcommand{\xr}{\xi_R}
\newcommand{\pall}{\parallel}
\newcommand{\el}{{\cal L}}
\newcommand{\er}{{\cal R}}

\begin{document}

\draft

\title{ Chiral Lagrangian with higher resonances \\
        and flavour $ SU(3) $ breaking }

\author{M. Hayakawa}
\address{ Department of Physics, Nagoya University,
          Nagoya 464-01, Japan }

\author{T. Kurimoto}
\address{ College of General Education,
          Osaka University, \\
          Toyonaka 560, Osaka, Japan }

%\author{T. Morozumi}
%\address{ Department of Physics, Hiroshima University, \\
%          Higashi-hiroshima 724, Japan }

\author{A. I. Sanda}
\address{ Department of Physics, Nagoya University,
          Nagoya 464-01, Japan }

\date{\today}

\maketitle

\abstract{
  A chiral Lagrangian
with $ SU(3) $ breaking and higher resonances
is proposed.
  In this model,
 the $ SU(3) $ breaking structure
 in vector-pseudoscalar sector is realized
 with the decay constants of pseudoscalar mesons
 and the masses of vector mesons used as inputs.
  We examine
 whether the resulting $ SU(3) $ breaking effect
 in the charge radii of pseudoscalar mesons
 is consistent with the experimental facts.
}

%\footnotetext[1]
% { electric address : hayakawa@eken.phys.nagoya-u.ac.jp }
%\footnotetext[2]
% { electric address : }
%\footnotetext[3]
% { electric address :
%   Morozumi@fusion.sci.hiroshima-u.ac.jp }
%\footnotetext[4]
% { electric address : sanda@eken.phys.nagoya-u.ac.jp }

\pacs{ }

\section{Introduction}
\label{sec:int}

 Quantitative prediction of rare $ K $ decay rates
based on the standard model is a necessary ingredient
in the program for understanding
the origin of $ CP $ violation.
 However, since the typical energy scale in $ K $ decays
$ \simeq M_K (\simeq 500\,{\rm MeV}) $
is too low for perturbative QCD to be applicable,
we need some method to estimate the long distance
contributions from QCD.
 The chiral symmetry imposes
strong constraints on low energy dynamics
associated with $ K $ decays
\cite{Ecker1,Bardeen,Buras,Ecker3,Kambor,Ecker4,Lim}.

\par
 In the usual formalism of chiral dynamics,
the energy dependence of the amplitudes involving
Goldstone bosons ( $ \pi, K $ )
is given as a power series expansion with respect to
the external momenta and quark masses.
 The coefficients of $ {\cal O}(p^4) $ terms
in the effective chiral Lagrangian
are explicitly determined from the experimental data
\cite{Gasser}.
 Later it has been recognized
that these coefficients of $ {\cal O}(p^4) $ terms
are saturated by the contributions from higher resonances
(vector, axial vector and scalar mesons)
\cite{Ecker3}.
 The importance of incorporating higher resonances
can also be seen in the result
that the $ \Delta I = \frac{1}{2} $ rule
is explained if we take
the propagation of scalar resonances
into account \cite{Lim}.

\par
 These results motivate us to the present work:
in this article we propose a model for
the chiral Lagrangian with scalar and vector mesons.

\par
 In QCD the $ SU(3) $ breaking effect is introduced
by the quark mass term
$ {\cal L}_m = - \bar{q}_L m^{(0)} q_R + h.c. $.
 In chiral dynamics, the scalar field plays the key role;
it develops the vacuum expectation value(VEV)
and induces the decay constants for pseudoscalar mesons.
 We shall see
that the explicit $ SU(3) $ breaking originating
from current quark masses
result in the differences among decay constants.
 We introduce vector meson
following the hidden local Lagrangian approach
\cite{Bando}.
The $ SU(3) $ breaking effects
in vector-pseudoscalar couplings
are generated by those of
decay constants of pseudoscalar mesons
and the masses of vector mesons.
 The explicit form of the potential terms
which gives the connection between the VEV's of scalars
and current quark masses
will be shown in Sec.\ref{sec:Preparation}.

\par
 Let us consider the expansion of the full Lagrangian
in powers of derivatives.
 At the order $ \del^2 $ level,
we have the kinetic term of pseudoscalars and
the mass term of vector mesons.
 There are an infinitely large number of terms
which are consistent
with chiral and hidden local symmetries.
 This is because arbitrary higher dimensional operators
of order $ \del^2 $ are allowed in our approach.
 In Sec.\ref{sec:o2} we argue first that
we can restrict ourselves to four-dimensional operators
since we are constructing
an effective Lagrangian which is valid below
the cutoff scale
$ \Lambda \simeq 4 \pi F_\pi \simeq 1\,{\rm GeV} $
\cite{Manohar}.

\par
 For a phenomenological test of this model
we calculate
the electromagnetic form factors for pseudoscalar mesons
and the $ K_{e3} $ form factor in Sec.\ref{sec:formfactor}.
 There,
the charge radii of $ \pi^+, K^+ $ and $ K^0 $
will be compared to their experimentally measured values.
 We will also find
that the normalization of $ K_{e3} $ form factor
$ F_+(0) \simeq 1 $ under fairly natural assumptions.
 In Sec.\ref{sec:extra},
we examine the possible effects
to charge radii
from the {\it extra} terms
which have not been included tentatively
in the reduced chiral Lagrangian.
 It is shown that these additional terms do not modify
the results obtained above.
 Sec.\ref{sec:discussion} is devoted
to discussion and summary.

%
%
%%%%%%%%%%%%%%%%%%%%%%%%%%%%%%%%%%%%%%%%%%%%%%%%%%%%%%%%
%                                                      %
%                       Lagrangian                     %
%                                                      %
%%%%%%%%%%%%%%%%%%%%%%%%%%%%%%%%%%%%%%%%%%%%%%%%%%%%%%%%
%
%
\section{Chiral Lagrangian}
\label{sec:Preparation}

\subsection{Preliminaries}

\par
 Let us begin our discussion
by recalling the way one derives
the chiral perturbation theory.

\par
 First we consider
$ U(N)_L \times U(N)_R $ linear $ \sigma $
model \cite{GellMann} which is defined by the Lagrangian
\cite{Ebert,anomaly}
\begin{eqnarray}
 {\cal L} &=&
 \displaystyle{
  \tr( \del_\mu M^\dagger \del^\mu M )
  + \mu^2 \tr( M^\dagger M )
  - \frac{\lambda_1}{4}
    \left\{ \tr( M^\dagger M ) \right\}^2
  - \frac{\lambda_2}{4} \tr( (M^\dagger M)^2 )
 } \nonumber \\
 && + \frac{1}{4 G_1}
    \tr \left( m^{(0)} ( M + M^\dagger ) \right),
\label{linear}
\end{eqnarray}
where $ m^{(0)} = {\rm diag}(m^{(0)}_u, m^{(0)}_d,\cdots) $
denotes current quark mass matrix,
and the last term breaks chiral symmetry explicitly.
 Using a unitary matrix $ \xi $
and a hermitian matrix $ \Sigma $,
we can decompose
$ N \times N $ complex matrix $ M $ in the form
\begin{eqnarray}
 M = \xi \Sigma \xi.
\nonumber
\end{eqnarray}
 The positivity of $ \mu^2 $ signals the occurrence of
spontaneous breakdown of chiral symmetry.
 This can be seen
by examining the vacuum expectation value(VEV)
$ \left< \Sigma \right> $
of $ \Sigma $ which is determined
by minimizing the potential
\begin{eqnarray}
 V(\left< \Sigma \right>) &=&
 \displaystyle{
  - \mu^2 \tr( \left< \Sigma \right>^2 )
  + \frac{\lambda_1}{4}
     \left\{ \tr(\left< \Sigma \right>^2) \right\}^2
  + \frac{\lambda_2}{4}
     \tr \left( \left< \Sigma \right>^4 \right)
 } \nonumber \\
 &&
 \displaystyle{
  - \frac{1}{2 G_1}
    \tr \left( m^{(0)} \left< \Sigma \right> \right)
 }.
\label{sigmapotential}
\end{eqnarray}
 Then the associated Goldstone bosons
$ \pi^a(a=0,1,\cdots N^2-1) $
are non-linearly realized in $ \xi $ as \cite{generator}
\begin{equation}
 \xi = \exp \left(
             \frac{i}{2} \sum_{a=0}^{N^2-1}
              \frac{\pi^a T^a}{F^a}
            \right),
\end{equation}
with dimension-one constants
$ F^a ( a = 1, \cdots, N^2 - 1 ) $.
\par
 The lowest order chiral Lagrangian used
in chiral perturbation theory can be obtained
in the infinitely heavy $ \Sigma^\prime $ limit,
where $ \Sigma^\prime = \Sigma - \left< \Sigma \right> $
is the dynamical degree of freedom associated with
the $ \Sigma $ field.
 Remember that the systematic chiral expansion
is a power series expansion
in the external momenta and quark masses.
 From the potential (\ref{sigmapotential})
we can see that
the splitting in the components of diagonal matrix
$ \left< \Sigma \right> $ is higher-order.
 Hence $ F^a = F_\pi ( a = 1, \cdots, N^2 - 1 ) $,
and by setting
$ \left< \Sigma \right> = \frac{F_\pi}{2} \bf{1} $
we obtain
\begin{eqnarray}
 {\cal L} = \frac{F_\pi^2}{4}
            \left\{
             \tr( \del_\mu U^\dagger \del^\mu U )
             + 2 B_0 \tr( m^{(0)} ( U + U^\dagger ) )
            \right\}
            + \ldots,
\nonumber
\end{eqnarray}
where $ U=\xi^2 $, $ B_0 = \frac{1}{4 G_1 F_\pi} $,
and the ellipsis contains
the higher-order terms in chiral expansion,
the irrelevant constant terms
and the terms involving physical scalar fields.

\par
 Now let the coupling $ \lambda_1 $ and $ \lambda_2 $
go to infinity with $ F_\pi $ kept fixed.
 Then scalar particles become so heavy
that they decouple effectively.
 Hence, in this limit, only Goldstone bosons survive
and their low energy behaviour is described
by the well-known leading order Lagrangian
\begin{eqnarray}
 {\cal L} = \frac{F_\pi^2}{4}
            \left\{
             \tr( \del_\mu U^\dagger \del^\mu U )
             + 2 B_0 \tr( m^{(0)} ( U + U^\dagger ) )
            \right\}.
\end{eqnarray}

\par
 Consider the case of two quark flavours( $ u $ and $ d $ )
and apply the corresponding chiral perturbation approach.
 Then the chiral expansion is in powers of external momenta
and the masses of these quarks
which are about $ 10\,{\rm MeV} $.
 Since the expansion parameter must be dimensionless,
we have to divide the typical momentum scale ($ p $)
and the masses of quarks by some constant $ \Lambda $
with mass dimension.
 This $ \Lambda $ is considered to be
the cutoff scale
under which our effective chiral description is valid,
and is about
$ 4 \pi F_\pi \simeq 1.2\,{\rm GeV} $ \cite{Manohar}
where $ F_\pi \simeq 93\,{\rm MeV} $
is the decay constant of $ \pi $ meson.
 Hence we expect, in this case,
that the convergence of this expansion
will be good as long as all external momenta are
sufficiently smaller than $ \Lambda $
( $ \frac{p^2}{\Lambda^2} \lesssim \frac{2m_u}{\Lambda}
    \simeq \times 10^{-2} $ ).

\par
 For the purpose of describing $ K $ mesons
within the framework of chiral dynamics,
we have to extend the above framework
such as to include strange quark.
 In this case one more expansion parameter
$ \frac{2 m_s}{\Lambda} = 0.18 \sim 0.36 $
for $ m_s = 100 \sim 200\,{\rm MeV} $
has to be taken into account.
 Also, in $ K $ decays,
the magnitudes of the typical momenta are
the order of $ K $ mass $ \simeq 500\,{\rm MeV} $.
 Thus we have
$ \left( \frac{p}{\Lambda} \right)^2
  \simeq \left( \frac{M_K}{\Lambda} \right)^2
  \simeq 0.21 $.
 Hence the expansion may not converge so rapidly as
in the case of $ SU(2)_L \times SU(2)_R $.

\par
 Let us give one such example.
 One of the successful challenge
is the explanation of the large enhancement
of $ \Delta I = \frac{1}{2} $ component in the
amplitude for $ K^0 \rightarrow 2\pi $
by incorporating scalar resonances
\cite{Lim}.
 There, the enhancement factor arises
from the scalar ( $ \sigma $ ) propagator
\cite{width}
\begin{eqnarray}
 \displaystyle{
  \frac{M_\sigma^2 - M_\pi^2}{M_\sigma^2 - M_K^2}
 }
 &=&
 \displaystyle{
  \frac{F_K}{3 F_\pi - 2 F_K}
  \simeq 2.2.
 }
\end{eqnarray}
 In the systematic expansion
of the chiral perturbation theory,
this factor will be obtained as the series
\begin{eqnarray}
 \displaystyle{
  \frac{F_K}{3 F_\pi - 2 F_K}
 }
 &=&
 \displaystyle{
  1 + \frac{3(F_K-F_\pi)}{F_K} + \cdots
  \cdots
 } \nonumber \\
 &=&
 \displaystyle{
  1 + 0.54 + (0.54)^2 + \cdots
 }. \nonumber
\end{eqnarray}
 The convergence of this series is very slow.
 Note that the $ SU(3) $ breaking gets amplified
when it comes into the numerator.
 Hence it is crucial to understand
how to include the $ SU(3) $ breaking.

\par
 To do this we must go back to
what we know about the low energy dynamics.
 One of the important facts is the vector meson dominance
structure which can be seen
in the pion-electromagnetic from
factor $ F_{\pi^+}(s) $ ( $ s $
is the invariant mass squared of virtual photon ).
 Under this hypothesis $ F_{\pi^+}(s) $ is given by
\begin{eqnarray}
 F_{\pi^+}(s) &=&
 \displaystyle{
  \frac{M_\rho^2}{M_\rho^2 - s}
  = 1 + \frac{1}{6} \left< r^2_{\pi^+} \right> s
    + \cdots
 },
\label{VD}
\end{eqnarray}
where $ M_\rho $ is the mass of $ \rho $ meson.
 Thus the charge radius $ \left< r^2 \right>_{\pi^+} $
is obtained as
\begin{eqnarray}
 \left< r^2 \right>_{\pi^+} &=&
 \displaystyle{
   6 \frac{1}{M_\rho^2}
   \simeq 0.41\,{\rm fm}^2,
 } \nonumber
\end{eqnarray}
which is in good agreement with its experimental value
$ 0.44\,{\rm fm}^2 $.
 It is also known
that the behaviour caused by the infinite series
in Eq.(\ref{VD}) recapitulates
the experimentally measured momentum dependence
of $ F_{\pi^+}(s) $.

\par
 In the systematic chiral expansion,
QCD dynamics appears
in the next-to-leading order terms,
i.e., $ {\cal O}(p^4) $ terms
which accompany
unknown constants $ L_1,\cdots ,L_{10} $ \cite{Gasser}.
 Each of these constants requires an experimental input.
 For example, $ L_9 $ has been determined
by using the charge radius
$ \left< r^2 \right>_{\pi^+} $ of $ \pi^+ $
as its input.
 However, as was shown before,
$ \left< r^2 \right>_{\pi^+} $ could be obtained
from the presence of $ \rho $ meson
with the vector meson dominance hypothesis.
 More generally,
$ L_1, \cdots, L_{10} $ are all determined from
the higher resonance contributions,
consistent with their proper values  \cite{Ecker3}.
 We shall, therefore, take an approach
in which
all higher-order effects in the chiral perturbation theory
can be reproduced by the introduction
of vector and scalar mesons \cite{Ecker3}.

%
%
%%%%%%%%%%%%%%%%%%%%%%%%%%%%%%%%%%%%%%%%%%%%%%%%%%%%%%%
%                                                     %
%                      Our model                      %
%                                                     %
%%%%%%%%%%%%%%%%%%%%%%%%%%%%%%%%%%%%%%%%%%%%%%%%%%%%%%%
%
%

\subsection{Model}
\label{subsec:model}

\par
 We propose a chiral Lagrangian
\cite{anomaly2}
based on the hidden local symmetry approach
\cite{Bando}
\begin{equation}
 {\cal L}_{\rm chiral} = {\cal L}_2 + {\cal L}_{\rm pot}
  - \frac{1}{2g_V^2} \tr( F_{V\mu\nu} F_V^{\mu\nu} ),
 \label{eqn:chiral}
\end{equation}
 where $ {\cal L}_2 $ consists of
the terms which contain two powers of derivatives.
The explicit form of which will be shown
in Sec.\ref{sec:o2}.
 $ F_V^{\mu\nu} $ is the field strength corresponding
to vector meson $ V_\mu $
\begin{eqnarray}
 F_V^{\mu\nu} &=& \del^\mu V^\nu - \del^\nu V^\mu
                - i\left[ V^\mu, V^\nu \right],
\nonumber
\end{eqnarray}
and $ g_V $ is the hidden local gauge coupling constant.
 $ {\cal L}_{\rm pot} $ determines
the VEV $ \VEV $ of
scalar nonet denoted by $ S $
and takes the form \cite{Pot}
\begin{eqnarray}
 {\cal L}_{\rm pot} &=&
  \mu^2 \tr(S^2)
   + \frac{1}{4G_1}
      \tr\left(
          S \left\{ \xi_L {\cal M} \xi_R^\dagger
                   + \xi_R {\cal M}^\dagger \xi_L^\dagger
            \right\}
         \right)
 \nonumber \\
% &&
%  + \tau\left\{
%            \det( S \xi_L {\cal M} \xi_R^\dagger)
%            + \det( S \xi_R {\cal M}^\dagger
%                    \xi_L^\dagger )
%           \right\}
% \nonumber \\
 &&
  - \frac{\lambda_1}{4} \left\{ \tr(S^2) \right\}^2
  - \frac{\lambda_2}{4}  \tr(S^4).
\label{potential}
\end{eqnarray}
 In this expression $ {\cal M} $ denotes
the $ 3 \times 3 $ matrix field
which couples to quarks in QCD Lagrangian \cite{Gasser}
as:
\begin{eqnarray}
 &&
  - \left( \bar{q}_L {\cal M} q_R
   + \bar{q}_R {\cal M}^\dagger q_L \right).
  \nonumber
\end{eqnarray}
 Hence, under chiral transformation
$ \{g_L,g_R\} \in
\left[ U(3)_L \times U(3)_R \right]_{\rm global} $,
$ {\cal M} $ is considered to ``transform''
as
\begin{eqnarray}
 {\cal M} &\rightarrow& g_L {\cal M} g_R^\dagger.
 \nonumber
\end{eqnarray}
 The vacuum configuration is determined through
Eq.(\ref{potential}) by setting  $ {\cal M} $
to be current quark mass matrix $ m^{(0)} $:
$ {\cal M} = m^{(0)} = {\rm diag}
  ( m^{(0)}_1, m^{(0)}_2, m^{(0)}_3 ) $.
 The differences in current quark masses
$ m^{(0)}_i(i=1,2,3) $
induce the splitting among the VEV's taken
by the neutral components of scalar $ S $.
 Throughout this paper
the isospin breaking effect due to quark masses
will be neglected:
$ m^{(0)}_1 = m^{(0)}_2 $.
 Hence the VEV( $ \Sigma^{(0)} $ )
of $ S $ takes the form
\begin{eqnarray}
 \VEV &=&
 \left(
  \begin{array}{ccc}
   \VEV_1 & & \\
    & \VEV_1 & \\
    & & \VEV_3
  \end{array}
 \right).
 \nonumber
\end{eqnarray}
 Since we do not perform an expansion
in powers of $ m^{(0)}_i $,
this splitting induces the differences
in the decay constants of pseudoscalar mesons
and the masses of vector mesons.

\par
 In the framework of hidden local symmetry,
$ \xi_L $ and $ \xi_R $ contain
not only the nonet of pseudoscalar mesons
but also a set of unphysical scalar degrees of freedom
($ \sigma $)
\begin{eqnarray}
 \xl &=& \xi(\sigma)\, \xi(\Pi)^\dagger,
   \nonumber \\
 \xr &=& \xi(\sigma)\, \xi(\Pi),
\end{eqnarray}
 where $ \xis = e^{i\sigma} $ and $ \xip = e^{i\Pi} $.
 The vector mesons are the gauge bosons
of hidden local symmetry
(in our case it is $ [U(3)_V]_{\rm local} $).
 The unphysical $ \sigma $ will be absorbed
by the vector mesons and give them nonzero masses
through Higgs mechanism \cite{Bando}.
 The transformation properties of $ \xi_L, \xi_R $
and $ S $ under chiral and hidden local symmetries are
\begin{eqnarray}
 \xl & \rightarrow & \xl^\prime = h\, \xl\, g_L^\dagger,
   \nonumber \\
 \xr & \rightarrow & \xr^\prime = h\, \xr\, g_R^\dagger,
   \nonumber \\
 S & \rightarrow & S^\prime = h\, S\, h^\dagger,
   \nonumber
\end{eqnarray}
where $ h $ is an element of
$ \left[ U(3)_V \right]_{\rm local} $.
 Then the Lagrangian $ {\cal L}_{\rm pot} $
in Eq.(\ref{potential})
can be seen to be invariant under
$ G \equiv \left[ U(3)_V \right]_{\rm local} \times
  \left[ U(3)_L \times U(3)_R \right]_{\rm global} $.
%except for the $ \tau $ term which breaks
%$ [U(1)_A]_{\rm global} $.
%
%

\par
 When we discuss the terms of $ {\cal O}(p^2) $
in Sec.\ref{sec:o2},
it is sufficient to use the following blocks,
each of which transforms as
$ A \rightarrow h\,A\,h^\dagger $ under $ G $;
\begin{eqnarray}
 \displaystyle{
  S,\ \alpha_{\pall\mu},\ \alpha_{\perp\mu},\
  D_\mu S,\ F_V^{\mu\nu},\ \frac{1}{G_1} \xl {\cal M}
  \xr^\dagger.
 }
\nonumber
\end{eqnarray}
 Here $ \alpha_{\pall\mu} $ and $ \alpha_{\perp\mu} $ are
\begin{eqnarray}
 \alpha_{\pall\mu} &=&
  \displaystyle{
   \frac{D_\mu \xl\,\xl^\dagger + D_\mu \xr\,\xr^\dagger}
    {2i}
  },
  \nonumber \\
 \alpha_{\perp\mu} &=&
  \displaystyle{
   \frac{D_\mu \xl\,\xl^\dagger - D_\mu \xr\,\xr^\dagger}
   {2i}
  }.
  \nonumber
\end{eqnarray}
 If we write the external vector fields as $ \el_\mu $
and $ \er_\mu $, associated with the gauging
of chiral symmetry $ [U(3)_L\times U(3)_R]_{\rm global} $,
then various covariant derivatives are given by
\begin{eqnarray}
 D_\mu \xl & = &
 \del_\mu \xl - i V_\mu \xl + i \xl \el_\mu,
  \nonumber \\
 D_\mu \xr & = &
 \del_\mu \xr - i V_\mu \xr + i \xr \er_\mu,
  \nonumber \\
 D_\mu S & = & \del_\mu S - i[V_\mu, S].
  \nonumber
\end{eqnarray}
  In constructing $ {\cal L}_2 $,
it is also necessary to know
the operation of charge conjugation($ {\cal C} $),
in particular,
on $ \alpha_\pall^\mu $ and $ \alpha_\perp^\mu $
\cite{Yamawaki}
\begin{eqnarray}
 {\cal C} &:&
 \alpha_\perp^\mu \rightarrow (\alpha_\perp^\mu)^T,\
 \alpha_\parallel^\mu
     \rightarrow -(\alpha_\parallel^\mu)^T.
\end{eqnarray}
 Hence the terms such as
$ {\tr}\left(
   \alpha_\parallel^\mu \left\{ S, D_\mu S \right\}
 \right) $
are forbidden by $ {\cal C} $ symmetry.
%
%
%
%
%
%%%%%%%%%%%%%%%%%%%%%%%%%%%%%%%%%%%%%%%%%%%%%%%%%%%%%%%%%%%
%                                                         %
%                       O(p^2) terms                      %
%                                                         %
%%%%%%%%%%%%%%%%%%%%%%%%%%%%%%%%%%%%%%%%%%%%%%%%%%%%%%%%%%%
%
%

\par
\section{$ {\cal O}(\lowercase{p}^2) $ terms}
\label{sec:o2}

 Here we discuss the $ {\cal O}(p^2) $ terms
with respect to the momentum expansion.
 The counting rule for momentum order is
\begin{itemize}
 \item $ \alpha_\parallel^\mu $ and $ \alpha_\perp^\mu $
       are $ {\cal O}(p) $.
 \item $ V_\mu, {\cal L}_\mu $
       and $ {\cal R}_\mu $ are $ {\cal O}(p) $.
       $ F_{\rm V}^{\mu\nu} $ is $ {\cal O}(p^2) $.
 \item $ S $ is $ {\cal O}(p^0) $.
 \item $ \frac{1}{G_1} {\cal M} $ is
       $ {\cal O}(p^2) $.
\end{itemize}
 We remark on the counting rule assigned
to the scalar field $ S $.
 Recall that
in the original hidden local symmetry approach
\cite{Bando}
the mass of vector meson is generated by the term
\begin{equation}
 a F_\pi^2 \tr(\alpha_\pall^\mu \alpha_{\pall\mu}),
\end{equation}
which is called $ {\cal O}(p^2) $ term.
 In our present scheme
the origin of the decay constants of pseudoscalars
is traced to the the nonzero VEV's of scalar $ S $.
 Since $ F_\pi $ is $ {\cal O}(p^0) $,
consistency will demand
that $ S $ is $ {\cal O}(p^0) $.
 Then $ {\cal O}(p^2) $ terms, such as
\begin{eqnarray}
 \tr( \{ S, \alpha_\pall^\mu \}
      \{ S, \alpha_{\pall\mu} \} ),
\end{eqnarray}
contribute to the masses for vector meson.

\par
 According to the above counting rule,
the fields which can participate
in $ {\cal O}(p^2) $ terms are
$ \alpha_\parallel^\mu, \alpha_\perp^\mu,
\frac{1}{G_1} {\cal M} $ and $ S $.
 The operator with dimension less than 4
requires compensation by the multiplication of constant(s)
with mass dimension when involved in the Lagrangian.
 In our scheme
we consider the constants with mass dimension
available to us are only
the VEV of $ S $ and $ \frac{1}{G_1} {\cal M} $.
 The latter appears only in the pseudoscalar mass terms
which have been already included in the potential term
in Eq.(\ref{potential}).
 Thus we do not need to be concerned with the operators
with dimension less than 4.

\par
 The operator with dimension( $ \equiv 4+d $) more than 4
has a coefficient of the form $ \frac{a}{\Lambda^d} $
where $ a $ is a constant of order 1,
and $ \Lambda $ is the cutoff
under which our effective chiral Lagrangian is expected
to describe the low energy behavior of QCD.
 Problematic thing is that
higher dimensional operators, such as
\begin{eqnarray}
 \frac{a}{\Lambda^d} \tr( S^d {\cal O}^{(4)} ),
\label{higher}
\end{eqnarray}
where $ {\cal O}^{(4)} $
is some four-dimensional and $ {\cal O}(p^2) $ operator,
can induce four-dimensional operators
by taking the VEV's of scalars.
 To see that we do not need to add these contributions,
write the operator in Eq.(\ref{higher}) as
\begin{eqnarray}
 &&
 \displaystyle{
  a
  \left\{
     \left( \frac{\VEV_1}{\Lambda} \right)^d
     \tr\,{\cal O}^{(4)}
   + \left(
       \left( \frac{\VEV_3}{\Lambda} \right)^d
       - \left( \frac{\VEV_1}{\Lambda} \right)^d
     \right)
     {\cal O}^{(4)}_{33}
  \right\}
 }.
\label{induced}
\end{eqnarray}

\par
 The first term in this expression can be absorbed
by the renormalization of the operator
$ \tr\,{\cal O}^{(4)} $.
 In the context where $ \Lambda \simeq 4 \pi F_\pi $
\cite{Manohar}, $ \VEV_1 \simeq \frac{F_\pi}{2} $.
 As will be shown later,
the splitting in the VEV's of $ S $ has the form:
$ \VEV_3 = \frac{M_\phi}{M_\rho} \VEV_1 $.
 Hence the second term in Eq.(\ref{induced})
is of order
\begin{eqnarray}
 \displaystyle{
  \left( \frac{\VEV_3}{\Lambda} \right)^d
  - \left( \frac{\VEV_1}{\Lambda} \right)^d
 }
 &\sim& 0.013\ \,(d=1),
 \nonumber \\
 &\sim& 0.0012 \,(d=2),
\end{eqnarray}
 which is smaller for larger $ d $.
 Hence, to order 1\% accuracy,
this second term can be dropped.
 Since we will not consider any processes
with scalar particles on the external lines,
this type of higher dimensional operators is not important.
 Therefore,
we can concentrate ourselves
with the dimension-four operators
which can be constructed from
$ \alpha_\parallel^\mu, \alpha_\perp^\mu $ and $ S $.
% Note that taking a determinant gives operators
%with dimension more than 4 at $ {\cal O}(p^2) $.
% Thus only the trace operation is allowed
%to make an invariant
%under chiral and hidden local transformations.

\par
 Below we list those operators that are Lorentz,
$ {\cal C} $,
$ {\cal P} $(Parity operation) invariant
and have chiral and hidden local symmetries:

\begin{equation}
 {\rm (Category\ 1)}
 \left\{
 \begin{array}{l}
  \tr([S,\alpha_\pall^\mu][S,\alpha_{\pall\mu}]) \\
  \tr(\{S,\alpha_\pall^\mu\}\{S,\alpha_{\pall\mu}\}) \\
  \tr([S,\alpha_\perp^\mu][S,\alpha_{\perp\mu}]) \\
  \tr(\{S,\alpha_\perp^\mu\}\{S,\alpha_{\perp\mu}\}) \\
  i\,\tr(\alpha_\pall^\mu\,[S,D_\mu S]) \\
  \tr(D^\mu S \,D_\mu S),
 \end{array}
 \right.
\label{o2operators1}
\end{equation}

\begin{equation}
 {\rm (C.2)}
 \left\{
 \begin{array}{l}
  \tr(S\,\alpha_\pall^\mu\,\alpha_{\pall\mu})\,\tr(S) \\
  \tr(S\,\alpha_\perp^\mu\,\alpha_{\perp\mu})\,\tr(S),
 \end{array}
 \right.
\end{equation}

\begin{equation}
 {\rm (C.3)}
 \left\{
 \begin{array}{l}
  \tr(\alpha_\pall^\mu\,\alpha_{\pall\mu})\,\tr(S)\,\tr(S)
   \\
  \tr(\alpha_\pall^\mu\,\alpha_{\pall\mu})\,\tr(S^2) \\
  \tr(\alpha_\perp^\mu\,\alpha_{\perp\mu})\,\tr(S)\,\tr(S)
   \\
  \tr(\alpha_\perp^\mu\,\alpha_{\perp\mu})\,\tr(S^2),
 \end{array}
 \right.
\end{equation}

\begin{equation}
 {\rm (C.4)}
 \left\{
 \begin{array}{l}
  \tr(S^2\,\alpha_\pall^\mu)\,\tr(\alpha_{\pall\mu}) \\
  \tr(S^2\,\alpha_\perp^\mu)\,\tr(\alpha_{\perp\mu}),
 \end{array}
 \right.
\end{equation}

\begin{equation}
 {\rm (C.5)}
 \left\{
  \begin{array}{l}
   \tr(S\,\alpha_\pall^\mu)\,\tr(S\,\alpha_{\pall\mu}) \\
   \tr(S\,\alpha_\perp^\mu)\,\tr(S\,\alpha_{\perp\mu}) \\
   \tr(D^\mu S)\,\tr(D_\mu S),
  \end{array}
 \right.
\end{equation}

\begin{equation}
 {\rm (C.6)}
 \left\{
  \begin{array}{l}
   \tr(S\,\alpha_\pall^\mu)\,\tr(\alpha_{\pall\mu})\,\tr(S)
    \\
   \tr(S\,\alpha_\perp^\mu)\,\tr(\alpha_{\perp\mu})\,\tr(S)
    ,
  \end{array}
 \right.
\end{equation}

\begin{equation}
 {\rm (C.7)}
 \left\{
  \begin{array}{l}
   \tr(\alpha_{\pall}^\mu)\,\tr(\alpha_{\pall\mu})\,
    \tr(S)\,\tr(S) \\
   \tr(\alpha_{\pall}^\mu)\,\tr(\alpha_{\pall\mu})\,
    \tr(S^2) \\
   \tr(\alpha_{\perp}^\mu)\,\tr(\alpha_{\perp\mu})\,
    \tr(S)\,\tr(S) \\
   \tr(\alpha_{\perp}^\mu)\,\tr(\alpha_{\perp\mu})\,
    \tr(S^2).
  \end{array}
 \right.
\label{o2operators2}
\end{equation}
 We divided operators into classes according to
the flavour $ SU(3) $ breaking structure
which will be generated
by the splitting among the VEV's of scalar $ S $.
 We first concentrate our attention on
the following {\it reduced} Lagrangian
which consists of the operators belonging
to the Category-1:
\begin{eqnarray}
 {\cal L}_2 &=&
 \displaystyle{
  \tr( \{S,\alpha_\perp^\mu\}\{S,\alpha_{\perp\mu}\} )
  + f\,
    \tr( [S,\alpha_\perp^\mu][S,\alpha_{\perp\mu}] )
 } \nonumber \\
 &&
 \displaystyle{
  + a\,
    \tr( \{S,\alpha_\pall^\mu\} \{S,\alpha_{\pall\mu}\})
  + b\,
    \tr( [S,\alpha_\pall^\mu][S,\alpha_{\pall\mu}])
 } \label{o2} \\
 && + d\,\tr( D^\mu S D_\mu S)
 + 2 c\,i\,\tr( \alpha_\pall^\mu\,[S,D_\mu S] ).
 \nonumber
\end{eqnarray}
 The effects of operators in the other categories
will be discussed later.

\par
 We first express fundamental quantities
in the pseudoscalar sector.
 Since isospin breaking is ignored here,
isospin nonsinglet mesons are always
in their mass eigenstates.
 We did not concern $ \eta $ and $ \eta^\prime $
explicitly here.
 So it is convenient to work in the nonet basis
\begin{eqnarray}
 \Pi &=& \frac{1}{\sqrt{2}}
 \left(
  \begin{array}{ccc}
   \displaystyle{
    \frac{\pi^0+\eta}{\sqrt{2}F_\pi}
   }
   &
   \displaystyle{
    \frac{\pi^+}{F_\pi}
   }
   &
   \displaystyle{
    \frac{K^+}{F_K}
   } \\
   \displaystyle{
    \frac{\pi^-}{F_\pi}
   }
   &
   \displaystyle{
    \frac{-\pi^0+\eta}{\sqrt{2}F_\pi}
   }
   &
   \displaystyle{
    \frac{K^0}{F_K}
   } \\
   \displaystyle{
    \frac{K^-}{F_K}
   }
   &
   \displaystyle{
    \frac{\bar{K}^0}{F_K}
   }
   &
   \displaystyle{
    \frac{\eta_{33}}{F_{33}}
   }
  \end{array}
 \right).
\nonumber
\end{eqnarray}
 Then we have
\begin{eqnarray}
 F_\pi &=& 2 \VEV_1,  \nonumber \\
 F_{33} &=& 2 \VEV_3,  \nonumber \\
 F_K^2 &=&
 \displaystyle{
  \frac{1}{4}
  \left\{
   (F_\pi+F_{33})^2 - f\, (F_{33}-F_\pi)^2
  \right\}
 }.
\label{decayconstant}
\end{eqnarray}
 The masses of $ \pi $ and $ K $ can be read off
from Eq.(\ref{potential})
\begin{eqnarray}
 M_\pi^2 &\equiv&
 \displaystyle{
  \frac{m^{(0)}_1 (2\VEV_1)}{2 G_1 F_\pi^2}
  = \frac{m^{(0)}_1}{2 G_1 F_\pi}
 },  \nonumber \\
 M_K^2 &\equiv&
 \displaystyle{
  \frac{(m^{(0)}_1+m^{(0)}_3)(\VEV_1+\VEV_3)}{4G_1 F_K^2}
 }
 =
 \displaystyle{
   \frac{(m^{(0)}_1+m^{(0)}_3)(F_\pi+F_{33})}
   {8 G_1 F_K^2}
 }.
\label{Pmass}
\end{eqnarray}

\par
 Next we turn our attention to the scalar-vector sector.
 There are transitions among
vector meson $ V_\mu $ , scalar $ \sigma $ and $ \Sigma $
where the field $ \Sigma $ is defined as
\begin{eqnarray}
 S &=&
 \displaystyle{
  \VEV + \frac{1}{\sqrt{d}} \Sigma
 }.
 \nonumber
\end{eqnarray}
 We define the unitary gauge as follows.
First we set $ \sigma = 0 $ and redefine the fields
$ V_\mu $ and $ \Sigma $ as
\begin{eqnarray}
 V_\mu &\rightarrow&
 \displaystyle{
  V_\mu + i \frac{1}{\sqrt{2}}
   \frac{C}{\sqrt{1-C^2}} \frac{g_V}{M_{K^*}} \del_\mu K^S,
 } \nonumber \\
 \Sigma_i^{\,j} &\rightarrow&
 \displaystyle{
  \frac{1}{\sqrt{1-C^2}} \Sigma_i^{\,j}
  \ {\rm for}\ (i,j) = (1,3),(2,3),(3,1),(3,2).
 }
\label{redef}
\end{eqnarray}
 In these expressions $ M_{K^*} $ is the mass of $ K^* $
(see Eq.(\ref{massofvector})),
and $ K^S $ consists of scalar components $ \kappa $
each of which has strange number
\begin{equation}
 K^S \equiv
 \left(
 \begin{array}{ccc}
  0 & 0 & -\kappa^+ \\
  0 & 0 & -\kappa^0 \\
  \kappa^- & \bar{\kappa}^0 & 0 \\
 \end{array}
 \right),
\end{equation}
and
\begin{eqnarray}
 C &\equiv&
 \displaystyle{
  \frac{d+c}{\sqrt{d}} g_V
  \frac{\VEV_3 - \VEV_1}{M_{K^*}}.
 }
\label{defofc}
\end{eqnarray}
 If we denote the vector meson matrix $ V_\mu $ as
\begin{equation}
 V_\mu = \frac{g_V}{\sqrt{2}}
 \left(
 \begin{array}{ccc}
  \frac{1}{\sqrt{2}} (\rho^0+\omega)_\mu & \rho^+_\mu
   & K^{*+}_\mu  \\
  \rho^-_\mu & \frac{1}{\sqrt{2}}(-\rho^0+\omega)_\mu
   & K^{*0}_\mu  \\
  K^{*-}_\mu & \bar{K}^{*0}_\mu & \phi_\mu
 \end{array}
 \right),
\end{equation}
 the masses of vector mesons are given by
\begin{eqnarray}
 M_\rho^2 &=&
 \displaystyle{
  a g_V^2 F_\pi^2 = M_\omega^2
 }, \nonumber \\
 M_\phi^2 &=&
 \displaystyle{
  a g_V^2 F_{33}^2
 }, \nonumber \\
 M_{K^*}^2 &=& \frac{a g_V^2}{4}
  \left\{
   (F_{33}+F_\pi)^2
   + \frac{d+2c-b}{a}(F_{33}-F_\pi)^2
  \right\}.
\label{massofvector}
\end{eqnarray}

\par
 We can see from Eq.(\ref{redef}) that
$ C^2 < 1 $ is necessary
for our model to become meaningful.
 Eq.(\ref{defofc}) tells us that
there is no mixing
between $ V_\mu $ and $ \Sigma $ when $ c = -d $.
 Note that if $ \frac{d+2c-b}{a} $ is order unity,
we have
\begin{eqnarray}
 M_{K^*} &=&
 \displaystyle{
  \frac{1}{2}
   (M_\rho+M_\phi),
 }
\label{vectorrelation}
\end{eqnarray}
since
\begin{equation}
 \frac{(F_{33} - F_\pi)^2}{(F_{33} + F_\pi)^2}
 =
 \frac{(M_\phi - M_\rho)^2}{(M_\phi + M_\rho)^2}
 \cong 0.02.
\end{equation}
 The $ f $ term
in the expression for $ F_K $ cannot be neglected:
in fact,
by considering $ f={\cal O}(1) $ and neglecting
the $ f $ term in Eq.(\ref{decayconstant}),
we obtain the relation
\begin{eqnarray}
 \frac{F_K}{F_\pi} &=&
 \frac{1}{2} \left( 1 + \frac{M_\phi}{M_\rho} \right)
 \simeq 1.16,
\end{eqnarray}
which is off by 6 \%.

\par
 We close this section by quoting the mass $ M_\kappa $
of $ \kappa $
since $ \kappa $ will contribute
to the $ K_{e3} $ form factor.
 $ M_\kappa $ is given by
\begin{equation}
 M_\kappa^2 =
 \displaystyle{
  \frac{1}{d}
  \frac{1}{1-C^2}
  \frac{1}{\VEV_3 - \VEV_1}
  \left(
   \frac{M_K^2 F_K^2}{\VEV_3 + \VEV_1}
   - \frac{M_\pi^2 F_\pi^2}{2 \VEV_1}
  \right)
 }.
\label{kappamass}
\end{equation}
 Using Eq.(\ref{decayconstant}),
it can also be expressed as
\begin{equation}
 \displaystyle{
 M_\kappa^2 =
 \frac{2}{d(1-C^2)}
 \left(
  \frac{2 F_K^2}{F_{33}^2 - F_\pi^2} M_K^2
  - \frac{1}
     {
       \displaystyle{
        \frac{F_{33}}{F_\pi}-1
       }
     } M_\pi^2
 \right)
 }.
\label{kappamass2}
\end{equation}
%
%
%
%
%
%%%%%%%%%%%%%%%%%%%%%%%%%%%%%%%%%%%%%%%%%%%%%%%%%%%
%                                                 %
%           Electromagnetic form factor           %
%           and K_{e3} form factor                %
%                                                 %
%%%%%%%%%%%%%%%%%%%%%%%%%%%%%%%%%%%%%%%%%%%%%%%%%%%
%
%
%
\section{ Electromagnetic form factors
          and $ K_{\lowercase{e}3} $ form factor }
\label{sec:formfactor}

 We now explore the electromagnetic form factors
and $ K_{e3} $ form factor based on
our chiral Lagrangian.
 In the standard model the external gauge fields
$ {\cal V}_\mu = \er_\mu + \el_\mu $
and $ {\cal A}_\mu = \er_\mu - \el_\mu $ take the form
\begin{eqnarray}
 {\cal V}_\mu &=&
 \displaystyle{
  \frac{2}{3} e \gamma_\mu
  \left(
  \begin{array}{ccc}
   2 &  0 &  0 \\
   0 & -1 &  0 \\
   0 &  0 & -1
  \end{array}
  \right)
 } \nonumber \\
 && +
 \displaystyle{
  \frac{g}{c_W} Z_\mu
  \left(
  \begin{array}{ccc}
   \frac{1}{2} - \frac{4}{3} s_W^2 & 0 & 0 \\
   0 & -\frac{1}{2} + \frac{2}{3} s_W^2 & 0 \\
   0 & 0 & -\frac{1}{2} + \frac{2}{3} s_W^2
  \end{array}
  \right)
 } \nonumber \\
 && +
 \displaystyle{
  \frac{g}{\sqrt{2}} W^+_\mu
  \left(
  \begin{array}{ccc}
   0 & c_1 & -s_1 c_3 \\
   0 &   0 & 0 \\
   0 &   0 & 0
  \end{array}
  \right)
  +
  \frac{g}{\sqrt{2}} W^-_\mu
  \left(
  \begin{array}{ccc}
   0 & 0 & 0 \\
   c_1 & 0 & 0 \\
   - s_1 c_3 & 0 & 0
  \end{array}
  \right)
 }, \nonumber \\
 {\cal A}_\mu
 &=&
 \displaystyle{
  -\frac{g}{2c_W} Z_\mu
  \left(
  \begin{array}{ccc}
   1 & 0 & 0 \\
   0 & -1 & 0 \\
   0 & 0 & -1
  \end{array}
  \right)
 } \nonumber \\
 &&
 \displaystyle{
  -
  \frac{g}{\sqrt{2}} W^+_\mu
  \left(
  \begin{array}{ccc}
   0 & c_1 & -s_1 c_3 \\
   0 & 0 & 0 \\
   0 & 0 & 0
  \end{array}
  \right)
  -
  \frac{g}{\sqrt{2}} W^-_\mu
  \left(
  \begin{array}{ccc}
   0 & 0 & 0 \\
   c_1 & 0 & 0 \\
   -s_1 c_3 & 0 & 0
  \end{array}
  \right)
 }, \nonumber
\end{eqnarray}
 where $ c_W = \cos\theta_W $
with the Weinberg angle $ \theta_W $,
and $ g $ is the $ SU(2) $ coupling constant.
 $ c_1 = \cos\theta_1, s_1 = \sin\theta_1,
c_3 = \cos\theta_3 $ and $ \theta_i $'s are the angles
parametrizing Kobayashi-Maskawa matrix \cite{Kobayashi}.

\par
 First we see that photon $ \gamma_\mu $
and neutral vector mesons mix
\begin{equation}
 {\cal L}_{\gamma-V} =
  -e \gamma^\mu \left[ g_\rho \rho^0_\mu
  + g_\omega \omega_\mu - g_\phi \phi_\mu \right],
\end{equation}
where \cite{Bando}
\begin{eqnarray}
 g_\rho &=& a\,g_V F_\pi^2 =
 \displaystyle{
  \frac{1}{g_V}\,M_\rho^2
 },  \nonumber \\
 g_\omega &=&
 \displaystyle{
  \frac{1}{3} g_\rho
 }, \nonumber \\
 g_\phi &=&
 \displaystyle{
  \frac{\sqrt{2}}{3} a\,g_V F_{33}^2
  = \frac{\sqrt{2}}{3} \frac{M_\phi^2}{g_V}
  = \frac{\sqrt{2}}{3} \frac{M_\phi^2}{M_\rho^2} g_\rho
 }.
\label{grhovmass}
\end{eqnarray}

\par
 The $ W-V $ mixing is given by
\begin{equation}
 {\cal L}_{W-V}^{\Delta S = 1 }
  = s_1 c_3 g g_{K^*} W^{-\mu} K^{*+}_\mu.
\end{equation}
 Here the constant $ g_{K^*} $ is calculated to be
\begin{eqnarray}
 g_{K^*} &\equiv&
 \displaystyle{
  \frac{a g_V}{8}
  \left\{
   ( F_{33} + F_\pi )^2
   - \frac{ - c + b }{a} ( F_{33} - F_\pi )^2
  \right\}
 }  \nonumber \\
 &=&
 \displaystyle{
  \frac{1}{2g_V}
  \left( 1 - \frac{C^2}{1+\bar{c}} \right) M_{K^*}^2
  = \frac{g_\rho}{2} \frac{M_{K^*}^2}{M_\rho^2}
  \left( 1 - \frac{C^2}{1+\bar{c}} \right)
 },
\label{gkstar}
\end{eqnarray}
 with $ \bar{c} \equiv \frac{c}{d} $.

\par
 The $ V-PP $ coupling($ P $ denotes pseudoscalar)
takes the form
\begin{eqnarray}
 {\cal L}_{V-PP} &=&
  -i \rho^{0\mu}
  \left(
   g_{\rho\pi\pi} \pi^+ \deldel_\mu \pi^-
   + g_{\rho KK} K^+ \deldel_\mu K^-
   - g_{\rho KK} K^0 \deldel_\mu \bar{K}^0
  \right)
  \nonumber \\
 &-& i g_{\omega KK} \omega^\mu
  \left(
  K^+ \deldel_\mu K^-
   - \deldel_\mu K^0 \bar{K}^0
  \right) \nonumber \\
 &+& i
  \displaystyle{
   \frac{g_{\phi KK}}{\sqrt{2}} \phi^\mu
   \left(
    K^+ \deldel_\mu K^- + K^0 \deldel_\mu \bar{K}^0
   \right)
  } \nonumber \\
 &-& i \displaystyle{\frac{g_{K^* K\pi}}{2} K^{*+\mu}}
  \left(
   K^- \deldel_\mu \pi^0
   - \sqrt{2} \pi^- \deldel_\mu \bar{K}^0
  \right)
  \nonumber \\
 &-& i
  \displaystyle{
   \frac{g_{K^* K\pi}}{\sqrt{2}} K^{*0\mu}
   \left(
    K^- \deldel_\mu \pi^+
    - \frac{1}{\sqrt{2}} \bar{K}^0 \deldel_\mu \pi^0
   \right)
  }
  \nonumber \\
 &+& \cdots,
\label{gVPP}
\end{eqnarray}
and each coefficient is given by KSRF(I) relation
\begin{eqnarray}
 g_{\rho\pi\pi} &=&
 \displaystyle{
   \frac{g_\rho}{2 F_\pi^2},
 } \nonumber \\
 g_{\rho KK} &=& g_{\omega KK} =
 \displaystyle{
  \frac{g_\rho}{4 F_K^2},
 } \nonumber \\
 g_{\phi KK} &=&
 \displaystyle{
  \frac{3 g_\phi}{2\sqrt{2} F_K^2},
 } \nonumber \\
 g_{K^* K\pi} &=&
 \displaystyle{
  \frac{g_{K^*}}{F_\pi F_K}.
 }
\label{KSRF}
\end{eqnarray}

\par
 We explore the numerical values
of various quantities by using the measured values of
$ F_\pi,F_K,M_\rho,M_\phi,M_{K^*} $
and $ g_{\rho\pi\pi} $.
 The results are shown in Table.\ref{tab:grho}
and \ref{tab:gvpp}.
 There, experimental uncertainties \cite{PDG}
$ \sqrt{2} F_\pi = 130.8\pm0.3\,{\rm MeV} $,
$ \sqrt{2} F_K   = 159.8\pm1.4\,{\rm MeV} $
and $ 5.9\le g_{\rho\pi\pi} \le 6.1 $
were taken into account.
 For the values of $ M_\rho $ and $ M_{K^*} $,
there is some theoretical uncertainties
due to our approximation.
 Here we used
$ (M_\rho)_{\rm \exp} \le M_\rho \le (M_\omega)_{\rm exp} $
for $ M_\rho $ and
$ (M_{K^{*+}})_{\rm exp} \le M_{K^*}
   \le (M_{K^{*0}})_{\rm exp} $ for $ M_{K^*} $.
 The value of $ g_{K^* K \pi} $ shown
in Table.\ref{tab:gvpp} was obtained
under the assumption that $ d+c $ is
at most of order unity:
 in fact, we have from Eq.(\ref{defofc})
\begin{eqnarray}
 C \simeq \frac{d+c}{\sqrt{d}} \times 0.10,
\nonumber
\end{eqnarray}
so that the quantity dependent on $ (d+c) $
in the parenthesis in Eq.(\ref{gkstar})
becomes $ (d+c) \times 0.01 $,
and this correction is within
the experimental errors accompanied with it.

\par
 The direct $ \gamma-PP $ and $ W-PP $ coupling are
\begin{eqnarray}
 {\cal L}_{\gamma PP} &=&
  -i e \gamma^\mu
  \left\{
   (1-r_\rho) (\pi^+ \del_\mu \pi^- - \del_\mu \pi^+ \pi^-)
  \right.
  \nonumber \\
 && \quad \quad \quad
 + (1-r_\phi) (K^+ \del_\mu K^- - \del_\mu K^+ K^-)
 \nonumber \\
 && \quad \quad \quad -
 \displaystyle{
 \left.
  r_\rho \frac{F_{33}^2-F_\pi^2}{3 F_K^2}
  (K^0 \del_\mu \bar{K}^0 - \del_\mu K^0 \bar{K}^0)
 \right\}
 }, \nonumber
\end{eqnarray}
\begin{eqnarray}
 {\cal L}^{\Delta S=1}_{W-PP} &=&
 \displaystyle{
  -i s_1 c_3 \frac{g}{4} W^{-\mu}
  \left\{
   \left( \frac{F_\pi}{F_K} - r_{K^*} \right)
   ( K^+ \del_\mu \pi^0 + \sqrt{2} \del_\mu \pi^+ K^0 )
  \right.
 } \nonumber \\
 && \quad\quad-
 \left.
 \displaystyle{
  \left( \frac{F_K}{F_\pi} - r_{K^*} \right)
  ( \del_\mu K^+ \pi^0 + \sqrt{2} \pi^+ \del_\mu K^0 )
 }
 \right\},
 \nonumber
\end{eqnarray}
 where $ r_\rho $ and $ r_\phi $ characterize
the vector dominance and are given by
\begin{eqnarray}
 r_\rho &\equiv&
 \displaystyle{
  \frac{a}{2} =
  \frac{2 g_{\rho\pi\pi}^2 F_\pi^2}{M_\rho^2}
 }, \nonumber \\
 r_\phi &=&
 \displaystyle{
  \frac{2 g_{\phi KK}^2 F_K^2}{M_\phi^2}
  \left( \frac{F_{33}^2 + 2 F_\pi^2}{3\,F_{33}^2} \right)
 }, \nonumber \\
 r_{K^*} &\equiv&
 \displaystyle{
  \frac{1}{2F_\pi F_K} \frac{a}{4}
  \left\{
   (F_{33}+F_\pi)^2 - \frac{b}{a}(F_{33}-F_\pi)^2
  \right\}
 } \nonumber \\
 &=&
 \displaystyle{
  \frac{2 g_{K^* K\pi}^2 F_\pi F_K}{M_{K^*}^2}
  \frac{
        \displaystyle{
         \left(
          1 - \frac{1+2\bar{c}}{(1+\bar{c})^2} C^2
         \right)
        }
       }
       {
        \displaystyle{
          \left( 1 - \frac{C^2}{1+\bar{c}} \right)^2
        }
       }
 }.
\label{rV}
\end{eqnarray}
 The expression for $ r_\rho $ has been obtained
before \cite{Bando}.
 This originates form the fact that
when ignoring the $ SU(3) $ breaking,
the first term and $ a $ term in Eq.(\ref{o2})
induces the same terms
as in the original hidden local Lagrangian
\cite{Bando}.
 However the expression for $ r_\phi $ is a peculiar one
arising from our model.
 Table.\ref{tab:vectordominance} shows
the numerical consequences
for these quantities in our model.
\par

 With these preparations, the electromagnetic form factors
can be obtained:
\begin{eqnarray}
 F_{\pi^+}(s) &=&
 \displaystyle{
  1 + r_\rho \frac{s}{M_\rho^2-s},
 } \nonumber \\
 F_{K^+}(s) &=&
 \displaystyle{
  \frac{1}{3}
  \left\{
   1 + 2 r_\rho \left( \frac{F_\pi}{F_K} \right)^2
       \frac{s}{M_\rho^2-s}
   + \bar{r}_\phi \frac{s}{M_\phi^2-s}
  \right\}
 }, \nonumber \\
 F_{K^0}(s) &=&
 \displaystyle{
  -\frac{1}{3}
  \left\{ r_\rho \left( \frac{F_\pi}{F_K} \right)^2
          \frac{s}{M_\rho^2-s}
   - \bar{r}_\phi
     \frac{s}{M_\phi^2-s}
  \right\}
 },
\label{electromagnetic}
\end{eqnarray}
 where $ s $ is
the invariant mass squared of virtual photon, and
\begin{eqnarray}
 \bar{r}_\phi &\equiv&
 \displaystyle{
  \frac{2 g_{\phi KK}^2 F_K^2}{M_\phi^2}
 }. \nonumber
\end{eqnarray}
 We can deduce the charge radius for each particle
\begin{eqnarray}
 \left< r^2 \right>_{\pi^+} &=&
 \displaystyle{
  6 \frac{r_\rho}{M_\rho^2}
 }, \nonumber \\
 \left< r^2 \right>_{K^+} &=&
 \displaystyle{
  6 \frac{\bar{r}_\phi}{M_\phi^2}
  = \left( \frac{F_K}{F_\pi} \right)^2
    \left< r^2 \right>_{\pi^+}
 }, \nonumber \\
 \left< r^2 \right>_{K^0} &=& 0.
\label{chargeradii}
\end{eqnarray}
 In deriving these expressions we use the relation
\begin{eqnarray}
 \bar{r}_\phi &=&
 \displaystyle{
  r_\rho \left( \frac{M_\phi}{M_\rho} \right)^2
  \left( \frac{F_\pi}{F_K} \right)^2
 }. \nonumber
\end{eqnarray}
 The numerical results for the charge radii are summarized
in Table.\ref{tab:chargeradii}.
 The $ SU(3) $ breaking in the charge radii appears
in the simple form
\begin{eqnarray}
 \displaystyle{
  \frac{ \left< r^2  \right>_{K^+} }
       { \left< r^2 \right>_{\pi^+} }
 }
 &=&
 \displaystyle{
  \left( \frac{F_\pi}{F_K} \right)^2
 }, \nonumber
\end{eqnarray}
the value of which is about 0.67
(its experimental value is 0.64)
so that in this aspect our model is not inconsistent with
the experiment.

\par
 Table.\ref{tab:chargeradii} shows
that our model gives slightly smaller values
for both $ \left< r^2 \right>_{\pi^+} $
and $ \left< r^2 \right>_{K^+} $.
 Also, as shown in Eq.(\ref{chargeradii}),
the charge radius $ \left< r^2 \right>_{K^0} $
for $ K^0 $ becomes exactly 0,
in contrast to its being experimentally small
but having a finite value
($ -0.054\pm 0.026 {\rm fm}^2 $).
 We will reexamine in Sec.\ref{sec:extra}
whether there is some modification to these results
by the addition of
$ 1/N_C $ non-leading but $ {\cal O}(p^2) $ terms,
which are not included in the reduced Lagrangian
in Eq.(\ref{o2}).
\par

 To calculate the $ K_{e3} $ form factor,
we further need
the transition term between $ \Sigma $ and $ W_\mu $,
and the interaction among
$ \kappa^- ,\pi^0 $ and $ K^+ $
\begin{eqnarray}
 {\cal L}_{S-W} &=&
 -i \sqrt{d(1-C^2)}
 \left( F_{33} - F_\pi \right)
 s_1 c_3 \frac{g}{4}
 W^+_\mu \del^\mu \kappa^- + (h.c.),
 \nonumber \\
 {\cal L}_{\kappa^- \pi^0 K^+} &=&
 \displaystyle{
  \frac{1}{\sqrt{d}} \frac{v_\sigma}{F_K}
  \kappa^- \del^\mu \pi^0 \del_\mu K^+
 }
 \displaystyle{
  - \frac{\lambda}{2} \frac{g_{K^* K\pi}}{M_{K^*}}
    \del^\mu \kappa^-
    \left( K^+ \deldel_\mu \pi^0 \right),
 } \nonumber
\end{eqnarray}
 where the constants $ v_\sigma $ and $ \lambda $ are
\begin{eqnarray}
 v_\sigma &\equiv&
 \displaystyle{
  \frac{1}{2\sqrt{1-C^2}}
  \left\{
   1 + \frac{F_{33} + F_\pi}{2 F_\pi}
       - f \frac{F_{33} - F_\pi}{2 F_\pi}
  \right\}
 } \nonumber \\
 &=&
 \displaystyle{
  \frac{1}{\sqrt{1-C^2}} \frac{F_K}{F_{33} - F_\pi}
  \left( \frac{F_K}{F_\pi} - \frac{F_\pi}{F_K} \right)
 }, \\
 \lambda &\equiv&
 \displaystyle{
  \frac{C \sqrt{1-C^2}}{1+\bar{c}-C^2}
 }. \nonumber
\end{eqnarray}
 The $ K_{e3} $ form factors
$ F_{\pm}(s) \left( s \equiv (p_K - p_\pi)^2 \right) $
is defined by
\begin{eqnarray}
 \left< \pi^0 (p_\pi) \right| \bar{s} (1-\gamma_5) u
 \left| K^+ (p_K) \right>
 &=& -
 \displaystyle{
  \frac{1}{\sqrt{2}} \left\{ F_+(s) (p_K+p_\pi)_\mu \right.
  \left. + F_-(s)(p_K-p_\pi)_\mu \right\}
 }. \nonumber
\end{eqnarray}
 Direct calculation shows that
$ F_+(s) $ and $ F_-(s) $ are given in our model by
\begin{eqnarray}
 F_+(s) &=&
 \displaystyle{
  \frac{1}{2} \left( \frac{F_K}{F_\pi} + \frac{F_\pi}{F_K}
              \right)
  - ( r_{K^*} - \bar{r}_{K^*} )
  + \bar{r}_{K^*} \frac{s}{M_{K^*}^2-s},
 } \nonumber \\
 F_-(s) &=&
 \displaystyle{
  \frac{1}{2} \left( \frac{F_K}{F_\pi} - \frac{F_\pi}{F_K}
              \right)
  - \bar{r}_{K^*}
            \frac{M_K^2-M_\pi^2}{M_{K^*}^2-s}
 } \label{Ke3} \\
 && \quad
 \displaystyle{
  + \frac{1}{M_{\kappa}^2-s}
    \left\{
     \frac{1}{2}
      \left( \frac{F_K}{F_\pi} - \frac{F_\pi}{F_K} \right)
      \left( M_K^2 + M_\pi^2 -s \right)
    \right.
 } \nonumber \\
 && \quad \quad \quad \quad \quad \quad \quad
  + \displaystyle{
     \left.
      \left( r_{K^*} - \bar{r}_{K^*} \right)
      \left( M_K^2 - M_\pi^2 \right)
     \right\}
    }. \nonumber
\end{eqnarray}
 Here $ \bar{r}_{K^*} $ is
\begin{eqnarray}
 \bar{r}_{K^*} &\equiv&
 \displaystyle{
  \frac{2 g_{K^* K\pi}^2 F_\pi F_K}{M_{K^*}^2}.
 }
\end{eqnarray}
 Hence the linear slope $ \lambda_{e3} $ defined by
\begin{eqnarray}
 F_+(s) &=&
 \displaystyle{
  F_+(0)
  \left(
   1 + \lambda_{e3} \frac{s}{M_\pi^2}
  \right)
  + {\cal O}\left(
             \left( \frac{s}{M_\pi^2} \right)^2
            \right)
 }, \nonumber
\end{eqnarray}
 takes the form
\begin{eqnarray}
 \lambda_{e3} &=&
 \displaystyle{
  \bar{r}_{K^*}
  \frac{M_\pi^2}{M_{K^*}^2 F_+(0)}
 },
 \label{lambda+}
\end{eqnarray}
 where
\begin{eqnarray}
 F_+(0) &=&
 \displaystyle{
  \frac{1}{2}
  \left( \frac{F_\pi}{F_K} + \frac{F_K}{F_\pi} \right)
  - ( r_{K^*} - \bar{r}_{K^*} )
 }, \nonumber \\
 r_{K^*} - \bar{r}_{K^*} &=&
 \displaystyle{
  \bar{r}_{K^*}
  \frac{C^2(1-C^2)}{(1+\bar{c}-C^2)^2}
  \simeq d\,\bar{r}_{K^*}\times 0.01
  \simeq d\times 0.01.
 }
\label{F+}
\end{eqnarray}
 In obtaining the first approximate relation
in Eq.(\ref{F+}), we made the same assumption
$ d+c \lesssim {\cal O}(1) $
as used for evaluating the numerical value
of $ g_{K^* K \pi} $.
 The second one follows by using the explicit value of
$ \bar{r}_{K^*} $ in Table.\ref{tab:vectordominance}.

\par
 There is still an ambiguity
due to the presence of factor containing $ d $.
 The value of $ d $ would be determined
if we knew the mass of strange scalar $ \kappa $
because from Eq.(\ref{kappamass2})
\begin{eqnarray}
 M_\kappa &\simeq&
 \displaystyle{
  \sqrt{\frac{2}{d}} \times 984\,{\rm MeV}.
 }
\label{d}
\end{eqnarray}
 However we only know that the strange scalar may be
rather heavy.
 We require that the mass of $ \kappa $
be greater than 980\,MeV.
 Also, following to the spirit of
the effective Lagrangian approach,
the mass of the particles contained explicitly
in the Lagrangian
should be less than the cutoff $ \Lambda $ of the theory
( $ \Lambda \simeq 4\pi F_\pi \simeq 1.2\,{\rm GeV} $ ).
 These considerations lead to
$ 980\,{\rm MeV} \lesssim M_\kappa
< \Lambda( \simeq 1.2\,{\rm GeV} ) $.
 This constraint to $ M_\kappa $ yields
$ r_{K^*} - \bar{r}_{K^*} \simeq 0.02 $
through the Eqs.(\ref{F+}) and (\ref{d}).
 Now we have
\begin{equation}
 F_+(0) \simeq
 \frac{1}{2}
 \left( \frac{F_K}{F_\pi} + \frac{F_\pi}{F_K} \right)
\end{equation}
 to the $ 2 $ \% level.
 Then the value of $ \lambda_{e3} $ is determined
which is shown in Table.\ref{tab:chargeradii}.
%
%
%
%
%%%%%%%%%%%%%%%%%%%%%%%%%%%%%%%%%%%%%%%%%%%%%%%%%%%
%                                                 %
%              Effects of extra terms             %
%                                                 %
%%%%%%%%%%%%%%%%%%%%%%%%%%%%%%%%%%%%%%%%%%%%%%%%%%%
%
%
%
%
\section{ Effects of extra terms }
\label{sec:extra}
 In this section, we examine the effects of the operators
in Eqs.$ (\ref{o2operators1})-(\ref{o2operators2}) $
to the the results obtained in the previous two sections.
 We make a few remarks
before adding these operators to
the reduced Lagrangian in Eq.(\ref{o2})
as {\it extra} terms.

\par
 First note that only operators in the Category 1
are allowed at the leading order of $ 1/N_C $ expansion
($ N_C $ is the number of colours)
if it is admitted as a proper approximation.
 This is because a trace in flavour space closes
the flow of flavour
so that we have one quark loop corresponding to one trace,
while the number of quark loops must be one
at the leading order of $ 1/N_C $ \cite{tHooft}.
 Hence at this order
only terms with just one trace are allowed.
 In this respect, the extra terms are non-leading.

\par
 The terms in which
two $ \alpha_\pall $'s are contained in separate traces
will induce the deviation
from $ SU(3) $ nonet basis for vector mesons.
 Since we know that
the ideal mixing of the octet and singlet vector mesons
is fairly good experimentally,
we do not consider such operators any more.
 On the other hand,
$ \tr(D_\mu S) \tr(D^\mu S)
( = \del_\mu \tr(S) \del^\mu \tr(S) ) $ and
the operators which contain two $ \alpha_\perp^\mu $'s
in different traces such as
$ \tr(S\alpha_\perp^\mu) \tr(S\alpha_{\perp\mu}) $
do not affect our interested quantities.
 Hence these operators will not be explicitly included
in the renewed Lagrangian.

\par
 Now we discuss the possible effects
to the charge radii of pseudoscalars
and the linear slope of $ K_{e3} $ form factor,
by the addition of
the following Lagrangian to the reduced Lagrangian
in Eq.(\ref{o2}):
\begin{eqnarray}
 {\cal L}_{\rm extra} &=&
 \dov \tr( S \alpha_\perp^\mu \alpha_{\perp\mu} )
 \tr(S)
 \nonumber \\
 && +
 \left\{
  \dlv \tr(S^2) + \Dlv \tr(S) \tr(S)
 \right\}
  \tr( \alpha_\perp^\mu \alpha_{\perp\mu} )
  \nonumber \\
 && +
 \dop \tr( S \alpha_\parallel^\mu \alpha_{\parallel\mu} )
 \tr(S) \nonumber \\
 && +
 \left\{
  \dlp \tr(S^2) + \Dlp \tr(S) \tr(S)
 \right\}
  \tr( \alpha_\parallel^\mu \alpha_{\parallel\mu} )
  .
\label{nextleading}
\end{eqnarray}
 Each coefficient denoted by $ \bar{\delta} $ includes
a possible suppression factor
associated with $ 1/N_C $ expansion.
 The similar remark is also made for $ \bar{\Delta} $,
but with a doubled suppression factor.

\par
 From the modified Lagrangian, Eq.((\ref{nextleading}),
\begin{eqnarray}
 F_\pi^2 &=&
 4 (\VEV_1)^2 + \dov \tr(\VEV) \cdot \VEV_1
 + T_\perp,
 \nonumber \\
 F_K^2 &=&
 \displaystyle{
  (\VEV_1+\VEV_3)^2 - f (\VEV_3-\VEV_1)^2
 } \nonumber \\
 && \quad
 \displaystyle{
  + \frac{1}{2} \dov \tr(\VEV) (\VEV_1+\VEV_3)
  + T_\perp,
 }
\end{eqnarray}
where $ T_\perp $ is defined as
\begin{equation}
 T_\perp \equiv
  \dlv \tr({\VEV}^2) + \Dlv (\tr(\VEV))^2.
\end{equation}
 The vector meson masses are also calculated to give
\begin{eqnarray}
 M_\rho^2 &=& M_\omega^2 =
 \displaystyle{
  g_V^2
  \left\{
   4a (\VEV_1)^2 + \dop \tr(\VEV)\cdot \VEV_1
   + T_\parallel
  \right\},
 } \nonumber \\
 M_{K^*}^2 &=&
 \displaystyle{
  g_V^2
  \left\{
   a (\VEV_3+\VEV_1)^2
   + (d+2c-b) (\VEV_3-\VEV_1)^2
  \right.
 } \\
 && \quad
 \displaystyle{
  \left.
   + \dop \tr(\VEV) \cdot \frac{\VEV_1+\VEV_3}{2}
   + T_\parallel
  \right\}
 }, \nonumber \\
 M_\phi^2 &=&
 \displaystyle{
   g_V^2
   \left\{
    4a (\VEV_3)^2 + \dop \tr(\VEV) \cdot \VEV_3
    + T_\parallel
   \right\},
 } \nonumber
\end{eqnarray}
where
\begin{equation}
 T_\pall \equiv
  \dlp \tr({\VEV}^2) + \Dlp (\tr(\VEV))^2.
\end{equation}
 If the large $ N_C $ argument can be applied,
the meaning of the VEV of scalar $ S $ remains unchanged
in this extension( $ \VEV_1 \simeq \frac{F_\pi}{2} $ ).
 Therefore the discussion
which asserts that higher dimensional operators are
less important than four-dimensional operators
in Sec.\ref{sec:o2}
makes sense even in this case.

\par
 By using these expressions we can directly check that
there is no modification to the expressions in
Eqs.(\ref{grhovmass}), (\ref{gkstar}),
(\ref{KSRF}), (\ref{rV}) and (\ref{electromagnetic})
by the presence of terms in Eq.(\ref{nextleading}).
 Therefore the charge radii are the same
as those obtained according to the reduced Lagrangian.
 We can also calculate the $ K_{e3} $ form factor
and find that the result is the same as in Eq.(\ref{Ke3}).
 Hence Eqs.(\ref{lambda+}) and (\ref{F+})
for the normalization and the linear slope of $ F_+(s) $
remain true.
 However, in this extended model,
we cannot express $ \kappa $ mass
in the form of Eqs.(\ref{kappamass2}) and (\ref{d})
( but Eq.(\ref{kappamass}) is true ).
 Since Eq.(\ref{F+}) holds even in this case,
$ F_+(0) \simeq 1 $ if we further assume that $ d $
is at most of order unity.

%
%
%
%
%%%%%%%%%%%%%%%%%%%%%%%%%%%%%%%%%%%%%%%%%%%%%%%%%%
%                                                %
%                 Discussion                     %
%                                                %
%%%%%%%%%%%%%%%%%%%%%%%%%%%%%%%%%%%%%%%%%%%%%%%%%%
%
%
%
\section{ Discussion and summary }
\label{sec:discussion}
 We have proposed a chiral Lagrangian
( Eqs.(\ref{eqn:chiral}) and (\ref{o2}))
with higher resonances (scalars and vectors),
paying a close attention
to the flavour $ SU(3) $ breaking structure
which shall be crucial
for the description of $ K $ decay.
 From the observation of
Eq.(\ref{vectorrelation})
 the resulting $ SU(3) $ breaking structure
in vector meson sector
seems to be well incorporated into our Lagrangian.
 Our model constructed here includes
not only $ {\cal O}(p^2) $ operators
but also the kinetic term of vector meson
which is $ {\cal O}(p^4) $
in the lowest order Lagrangian.
 Hence the quantities which requires higher-order terms
can be calculated.
 Our challenge with the use of our model
is to ask whether it can give sufficient predictions
consistent with the experimental facts
only by taking the flavour $ SU(3) $ breaking structure
into account, with only such an $ {\cal O}(p^4) $ term.

\par
 The first test of our model was performed
by confronting its prediction
for charge radii of pseudoscalars
$ \pi^+, K^+ $ and $ K^0 $
with their experimentally obtained values.
 As a result, the predicted value for
$ \left< r^2 \right>_{\pi^+} $
is found to be slightly smaller.
 Also the charge radius of $ K^0 $ becomes exactly zero.

\par
 These consequences does not change even if we add
the extra terms in Eq.(\ref{nextleading})
as was shown in Sec.\ref{sec:extra}.
 From this fact we can say that
the $ SU(3) $ breaking structure
in the pseudoscalar-vector sector in our model
are determined
only by the chiral and hidden local symmetries.

\par
 We finally remark on the application of our Lagrangian
to the actual calculation of $ K $ decays.
 The most familiar framework for it
will be the effective Hamiltonian method \cite{Gilman}
with the factorization hypothesis.
 There the long distance contribution from QCD
is considered to reside in the hadronic matrix elements of
four-Fermi operators and is expected to be calculated
by using our chiral Lagrangian.
 The unreliability to the results obtained
by following this approach
comes from the strong dependence of them
on the renormalization point
which is conceptually the matching scale
between short and long distance physics \cite{Buras}.
 Hence we must at first reexamine this point
in order to give definite prediction
form our chiral Lagrangian.
\acknowledgments{
 We thank K.Yamawaki
for helpful discussions on hidden local symmetry.
}

\vfill
\newpage
%
%
%%%%%%%%%%%%%%%%%%%%%%%%%%%%%%%%%%%%%%%%%%%%%%%%%%%
%                                                 %
%               g_V(V=\rho,\omega,\phi)           %
%                                                 %
%%%%%%%%%%%%%%%%%%%%%%%%%%%%%%%%%%%%%%%%%%%%%%%%%%%

\begin{table}
\caption[ ]{  $ g_V(V=\rho,\omega,\phi) $ coupling
              with $ g_{\rho\pi\pi} $
              used as input through KSRF(I)
              relation(in unit[$ {\rm GeV}^2 $])
           }
\label{tab:grho}
\begin{center}

\begin{tabular}{cll}
  & Prediction & Experiment \\
\hline
 $ g_\rho $   & $ 0.103\pm0.003 $ & $ 0.117\pm0.003 $ \\
 $ g_\omega $ & $ 0.034\pm0.001 $ & $ 0.036\pm0.001 $ \\
 $ g_\phi $   & $ 0.084\pm0.003 $ & $ 0.081\pm0.002 $
\end{tabular}

\end{center}

\end{table}

%%%%%%%%%%%%%%%%%%%%%%%%%%%%%%%%%%%%%%%%%%%%%%%%%%%%%
%                                                   %
%                   g_{VPP} coupling                %
%                                                   %
%%%%%%%%%%%%%%%%%%%%%%%%%%%%%%%%%%%%%%%%%%%%%%%%%%%%%

\begin{table}

\caption[ ]{ $ g_{VPP} $ coupling
             with $ g_{\rho\pi\pi} $ used as input }
\label{tab:gvpp}

\begin{center}

\begin{tabular}{cll}
  & Prediction & Experiment \\
\hline
 $ g_{\rho\pi\pi} $   & $ 6.00\pm0.10({\rm input}) $
  & $ 6.00\pm0.10 $ \\
 $ g_{\phi KK} $ & $ 6.98\pm0.39 $
  & $ 6.58\pm0.22 $ \\
 $ g_{K^* K\pi} $ & $ 6.55\pm0.34 $ & $ 6.47\pm0.11 $
\end{tabular}

\end{center}

\end{table}

%%%%%%%%%%%%%%%%%%%%%%%%%%%%%%%%%%%%%%%%%%%%%%%%%%%%%%%
%                                                     %
%                     r_\rho,r_\phi                   %
%                                                     %
%%%%%%%%%%%%%%%%%%%%%%%%%%%%%%%%%%%%%%%%%%%%%%%%%%%%%%%

{
\narrowtext

\begin{table}

\caption[ ]{ $ r_\rho, r_\phi $ and $ {\bar r}_{K^*} $ }
\label{tab:vectordominance}

\begin{center}

\begin{tabular}{cl}
 $ r_\rho $ & $ 1.028\pm0.058 $ \\
 $ r_\phi $ & $ 0.859\pm0.074 $ \\
 $ \bar{r}_{K^*} $ & $ 1.069\pm0.048 $
\end{tabular}

\end{center}

\end{table}
}

%%%%%%%%%%%%%%%%%%%%%%%%%%%%%%%%%%%%%%%%%%%%%%%%%%%%
%                                                  %
%                   charge radii                   %
%                                                  %
%%%%%%%%%%%%%%%%%%%%%%%%%%%%%%%%%%%%%%%%%%%%%%%%%%%%

\begin{table}

\caption[ ]{
             charge radii of pseudoscalars
             and $ \lambda_{e3} $
             (charge radii are in unit
              [$ {\rm fm}^2 $].).
           }
\label{tab:chargeradii}

\begin{center}

\begin{tabular}{cll}
  & Prediction & Experiment \\
\hline
 $ \left< r^2 \right>_{\pi^+} $
  & $ 0.401\pm0.030 $ & $ 0.439\pm0.008 $ \\
 $ \left< r^2 \right>_{K^+} $ & $ 0.269\pm0.026 $
  & $ 0.28\pm0.07 $ \\
 $ \left< r^2 \right>_{K^0} $
  & $ 0.00 $ & $ -0.054\pm0.026 $ \\
 $ \lambda_{e3} $ & $ 0.026\pm0.004 $
  & $ 0.0286\pm0.0022 $
\end{tabular}

\end{center}

\end{table}


\begin{references}
\bibitem{width}
  The argument holds
  when we introduce the width for the $ \sigma $ particle.
\bibitem{generator}
  $ T^a ( a = 1, \cdots N^2-1) $ are the generators of
  the fundamental representation of $ SU(N) $,
  and $ T^0 $ denotes $ U(1) $ generator;
  $ T^0 = \frac{1}{\sqrt{2N}} \left(1,\cdots,1 \right) $.
  All these matrices should be normalized
  such that $ \tr( T^a T^b ) = \frac{1}{2} \delta^{ab}
  (a,b=0,1,\cdots N^2-1) $.
\bibitem{anomaly2}
  When we consider the process which occurs
  through the existence of chiral anomalies,
  we add the appropriate anomaly terms.
\bibitem{Pot}
   From the symmetry argument, we have more terms
  for the quartic coupling of S
  ( $ \tr(S) \tr(S^3) $,
  $ \left\{ \tr(S) \right\}^2 \tr(S^2) $ and
  $ \left\{ \tr(S) \right\}^4 $ ).
   Even if they are included in $ {\cal L}_{\rm pot} $,
  all the results to be obtained in this paper
  do not change.
   Thus we will not be concerned with this matter here.
\end{references}
\end{document}